\documentclass[a4paper,11pt]{article}

\usepackage{pos}
\usepackage{bm}
\usepackage{comment}
\usepackage{mathtools}
\usepackage{lipsum} 
\usepackage{siunitx}
\usepackage{lineno}
\usepackage{caption}
\usepackage{subcaption}

\title{Conditional normalizing flows for IceCube event reconstruction}

\ShortTitle{Conditional normalizing flows for IceCube event reconstruction}

\author{The IceCube Collaboration \\{\normalsize \normalfont(a complete list of authors can be found at the end of the proceedings)}\\}

\emailAdd{thorsten.gluesenkamp@physics.uu.se}

\abstract{

The IceCube Neutrino Observatory is a cubic-kilometer high-energy neutrino detector deployed in the Antarctic ice. Two major event classes are charged-current electron and muon neutrino interactions. In this contribution, we discuss the inference of direction and energy for these classes using conditional normalizing flows. They allow to derive a posterior distribution for each individual event based on the raw data that can include systematic uncertainties, which makes them very promising for next-generation reconstructions. 

For each normalizing flow we use the differential entropy and the KL-divergence to its maximum entropy approximation to interpret the results.
The normalizing flows correctly incorporate complex optical properties of the Antarctic ice and their relation to the embedded detector. For showers, the differential entropy increases in regions of high photon absorption and decreases in clear ice. For muons, the differential entropy strongly correlates with the contained track length. Coverage is maintained, even for low photon counts and highly asymmetrical contour shapes. For high-photon counts, the distributions get narrower and become more symmetrical, as expected from the asymptotic theorem of Bernstein-von-Mises. For shower directional reconstruction, we find the region between 1 TeV and 100 TeV to potentially benefit the most from normalizing flows because of azimuth-zenith asymmetries which have been neglected in previous analyses by assuming symmetrical contours. Events in this energy range play a vital role in the recent discovery of the galactic plane diffuse neutrino emission.

\vspace{4mm}
{\bfseries Corresponding authors:}
Thorsten Gl\"usenkamp$^{1*}$\\
{$^{1}$ \itshape Uppsala University, Uppsala, Sweden}\\[4mm]
$^*$ Presenter

\ConferenceLogo{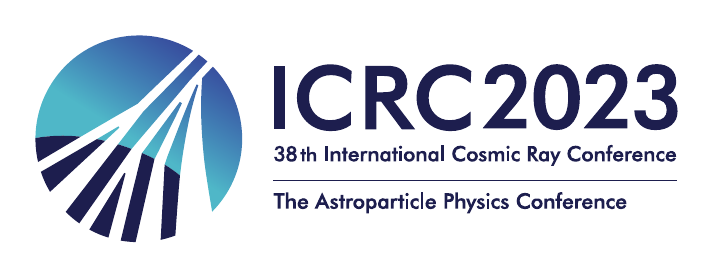}

\FullConference{The 38th International Cosmic Ray Conference (ICRC2023)\\ 26 July -- 3 August, 2023\\ Nagoya, Japan}
}

\begin{document}

\maketitle

\section{Introduction}\label{sec:intro}

High-energy neutrino astronomy emerged in the last decade, starting with the discovery of a diffuse neutrino flux in 2013 \cite{hese_refined}. In 2017, the first evidence for a neutrino source, the blazar \texttt{TXS 0506+056}, was found using a FERMI-LAT gamma-ray follow-up of a neutrino alert and historical analysis of neutrino data \cite{txs}. In 2020, the point source search analysis improved by modeling the signal probability density function based on reconstructed quantities \cite{ps_improvements}, which solved previous shortcomings in per-event contour coverage. It led to the first time-independent discovery of a neutrino source, the Seyfert galaxy \texttt{NGC 1068}. This new approach allowed for unbiased estimation of point source spectral parameters but sacrificed some sensitivity by averaging over events. Deep learning techniques, specifically conditional normalizing flows, now offer the potential to model non-symmetrical per-event uncertainties with correct coverage and improve point source capabilities yet again. Some initial studies regarding their performance for electron and muon neutrino charged current interactions are discussed in the following.

\section{Conditional normalizing flows}\label{sec:nfs}
Normalizing flows define flexible $d$-dimensional probability density functions and are defined with a change of basis formula via
\begin{equation}
p_{\theta}(x) = p_0(f_{\theta}^{-1}(x)) \cdot |\mathrm{det}J_{\theta}^{-1}(x)|, \label{eq:nf}
\end{equation}
where $x$ is a $d$-dimensional vector, $f_{\theta}$ a bijective and differentiable function with parameters $\theta$ and $J$ is the Jacobian of $f$.
The base distribution $p_0$ is typically chosen to be a standard normal distribution in $d$ dimensions, which has certain advantages for coverage calculations \cite{cov_paper}. The parameters $\theta$ that define the flow-defining function $f$ are the actual parameters of the PDF that is created in this way.
One can introduce conditional dependency into the PDF by letting the flow parameters $\theta$ be the output of a neural network:
\begin{equation}
p_{\theta}(x|y) = p_0(f_{g_\theta(y)}^{-1}(x)) \cdot |\mathrm{det}J_{g_\theta(y)}^{-1}(x)| \label{eq:conditional_nf}
\end{equation} 
The function $g$ represents a neural network with parameters $\theta$, which now are equivalent to the parameters that define the conditional PDF.
Normalizing flows can be defined over Euclidean space, but also over manifolds like spheres \cite{nf_spheres}. For this contribution, we utilize three types of conditional normalizing flows. For the neutrino energy we use 1-d Euclidean flows consisting of two Gaussianization flow units \cite{gaussianization_flows} ($f_{G}$) and an affine flow ($f_{A}$). The target space is $\mathrm{log}_{10}(E_\nu)$, i.e. the logarithm of the neutrino energy. For the neutrino direction we use 
 2-d spherical normalizing flows consisting of exponential-map flows with exponentiated potential \cite{nf_spheres} ($f_E$) and rotations ($f_R$). The exact definition of the flow-defining functions are given in table \ref{tab:flow_defs}. All normalizing flows have been implemented in the open-source package \texttt{jammy\_flows} \cite{jf} which has been used throughout this work. We use the coverage calculation for spherical flows as introduced in \cite{cov_paper}.
\begin{table}[h!]
\begin{center}
\begin{tabular}{||c c||} 
 \hline
 type & flow function  \\ [0.5ex] 
 \hline\hline
 1-d Euclidean & $f(z)=[f_{A} \circ f_{G} \circ f_{G}](z)$ \\ 
 \hline
 2-d spherical & $f(z)=[ (f_{R} \circ f_{E})^8](z)$  \\
 \hline
\end{tabular}
\caption{Flow definitions used for individual 1-d Euclidean and 2-d spherical flows. Iterative nesting of the bijective flow functions is indicated by $\circ$. }
\label{tab:flow_defs}
\end{center}
\end{table}

The neural network $g_\theta(x)$ (see eq. \ref{eq:conditional_nf}) is a graph neural network with base architecture as defined in \cite{gnn_paper}, whose output is mapped via a multilayer-perceptron (MLP) to all the flow parameters. The difference with respect to the  architecture in \cite {gnn_paper} is that the input features are not individual photon hit information, but a list of summary statistics of the noise-cleaned photon hits in a given optical module. Furthermore we use different aggregation functions and non-linearities.

\section{Reconstructions of IceCube events}\label{sec:reconstructions}

\begin{figure*}[hbt]
\centering
\begin{subfigure}[b]{0.49\textwidth}
         \centering
         \includegraphics[width=\textwidth]{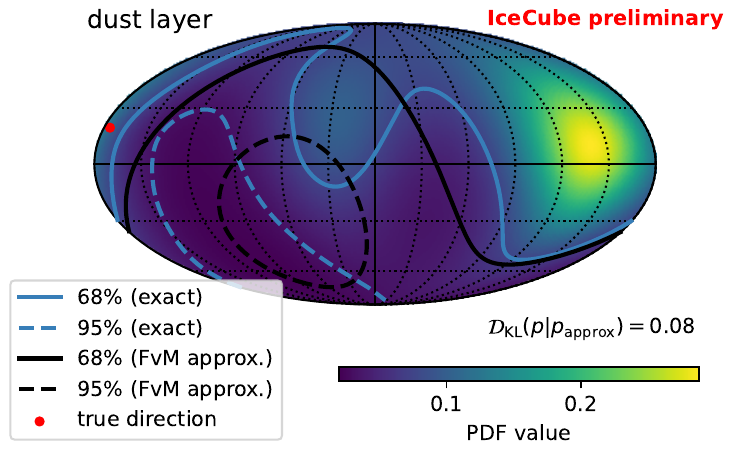}
         \caption{example of direction posterior in the dust layer}
         \label{fig:dust_layer_example}
\end{subfigure}
\begin{subfigure}[b]{0.49\textwidth}
         \centering
         \includegraphics[width=\textwidth]{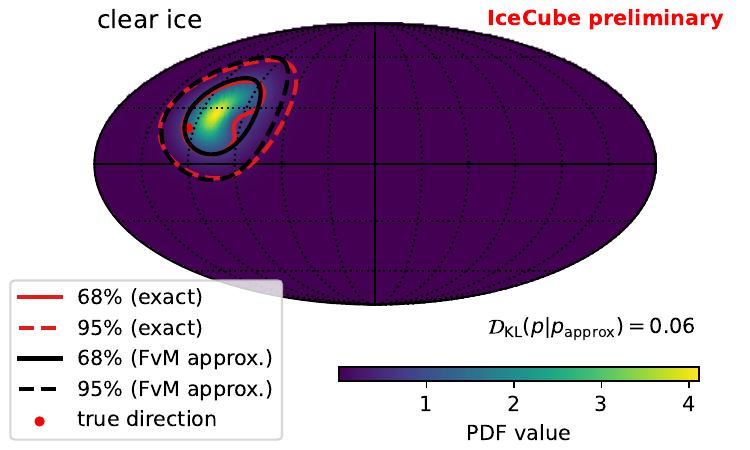}
         \caption{example of direction posterior in clear ice}
         \label{fig:clear_ice_example}
\end{subfigure}
\begin{subfigure}[b]{0.49\textwidth}
         \centering
         \includegraphics[width=\textwidth]{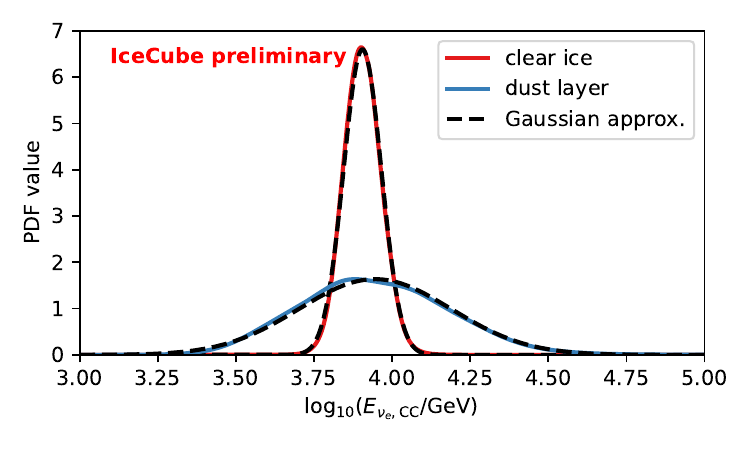}
         \caption{energy posteriors of events from (a) and (b)}
         \label{fig:energy_posterior_examples}
\end{subfigure}
\begin{subfigure}[b]{0.49\textwidth}
         \centering
         \includegraphics[width=\textwidth]{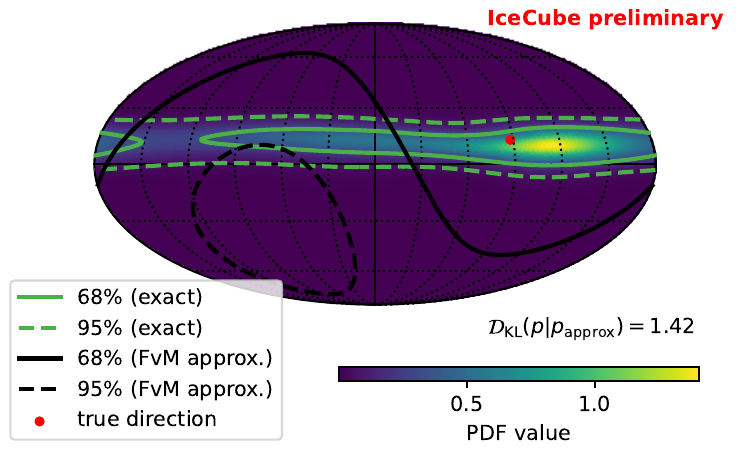}
         \caption{direction posterior of a high KL-divergence event}
         \label{fig:high_kl_example}
\end{subfigure}

\caption{Visualizations of directional and energy posterior regions for two $\nu_e$ charged-current events with about $10 \ \mathrm{TeV}$ neutrino energy. The event shown in a) is located in the dust layer, while the event shown in b) is located in a clearer ice region. The corresponding energy posteriors are shown in (c). The true values for the directions are indicated as red dots. Also shown is an example with high KL-divergence (d) as discussed in section \ref{sec:electron_dir}. The respective Fisher-von-Mises and Gaussian approximations are depicted in black or black-dashed. }
\label{fig:example_contours}
\end{figure*}

\textbf{\textit{Overview of training and coverage:}}
We performed three neural-network training runs. Two of them were trained on $\approx 1.2$ million events of electron-neutrino charged-current interactions to reconstruct a posterior over the neutrino energy (1d) or direction (2d). The third was trained on $\approx 1.4$ million events of muon neutrino charged-current interactions to reconstruct a posterior over direction. The energy spectrum during training had a spectral index of $-1.8$ to obtain an approximately flat distribution after the detector response. The following visualizations and evaluations are performed on independent test datasets with $\approx 100000$ events each. Example posteriors for two events with $\approx 10 \ \mathrm{TeV}$ are shown in fig. \ref{fig:example_contours}.
The first is a shower event in the "dust layer" (fig. \ref{fig:dust_layer_example}), a region roughly 2.1 km below the surface with a high dust concentration and therefore high photon absorption. Because of low optical transparency, the posterior region is spread out. The second is a shower event  in the "clear ice" region (fig. \ref{fig:clear_ice_example}) situated between 2.3-2.4 km below the surface with particularly low dust contamination. The respective energy posteriors of the same events are approximately Gaussian (fig. \ref{fig:energy_posterior_examples}), but the dust layer event has a significantly larger uncertainty.  In both example posteriors for the direction, the true direction is indicated by a red dot and lies in the respective $68 \%$ interval.

\begin{figure*}[hbt]
\centering
\includegraphics[width=0.99\textwidth]{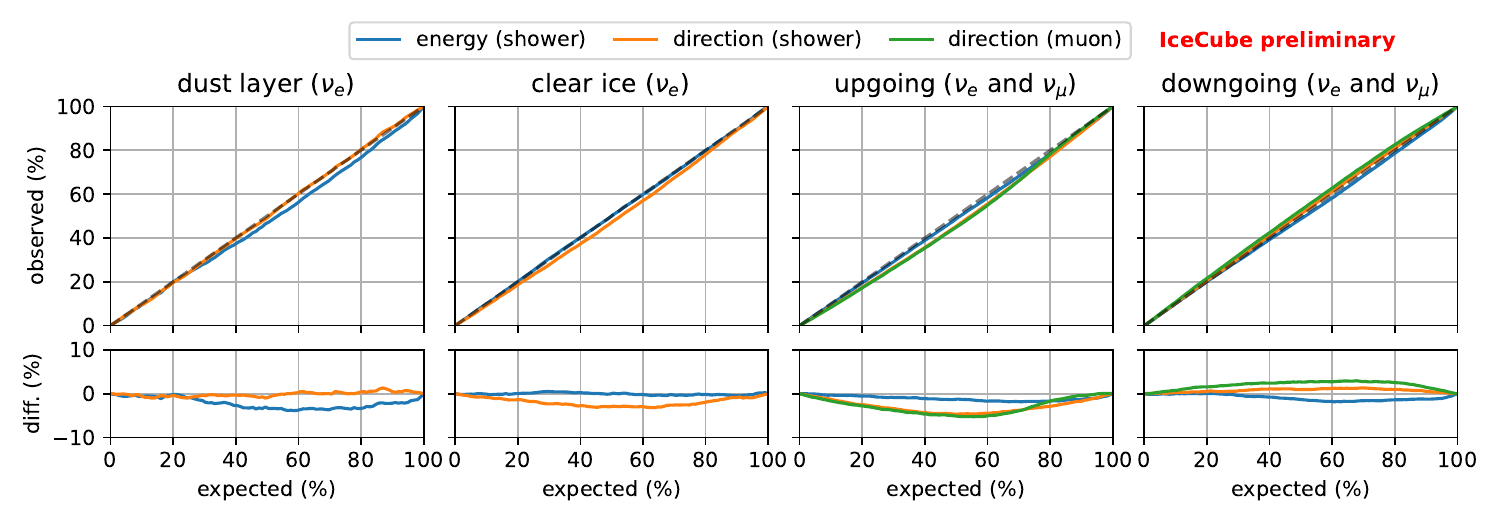}
\caption{Coverage (explained in the text) for electron neutrino direction, electron neutrino energy, and muon direction PDFs for four different types of events. The events in the dust layer and clear ice region are selected based on the depth of the neutrino interaction, and are only shown for electron neutrinos.}
\label{fig:coverage}
\end{figure*}

We used the methodology to calculate coverage for normalizing flows introduced in \cite{cov_paper} and apply it on the different topologies. Coverage indicates how often a true value falls within a certain expected probability contour, which is what we call "observed" coverage compared to the "expected" contour containment. Fig. \ref{fig:coverage} shows the coverage behavior for shower direction, shower energy, and muon direction - split up into different categories. The muon direction coverage is only shown for the up- and downgoing sets, as the depth of the interaction position is not meaningful for arbitrarily aligned tracks. Coverage is within a few percent, even for asymmetric contour shapes.

\textbf{\textit{Quantifying non-symmetrical PDF shapes with information theory:}}
In the following we calculate symmetric maximum entropy approximations to the normalizing-flow distributions. For the Euclidean normalizing flow, the maximum-entropy approximation is a Gaussian distribution. For the spherical normalizing flow, the symmetric maximum-entropy approximation corresponds to a Fisher-von-Mises (FvM) distribution. We then calculate two information-theoretic quantities to draw conclusions about each individual event: the differential entropy for a performance measure and the KL-divergence between the exact distribution and the approximation for a measure of asymmetrical contour shapes.

A useful quantity to gauge the reconstruction performance for non-symmetric distributions is the differential entropy $S_d$. It is defined as 
\begin{equation}
    S_{d}[p(x)]=- \int p(x) \ \mathrm{ln}(p(x)) \ \mathrm{dx},
\end{equation}
and measures how spread out the distribution is. To first order, it is useful to think about it as being proportional to the logarithm of the variance, as we have the closed-form solutions
\begin{align}
    S_{d}[\mathcal{G}_{1d}] &= \frac{1}{2}+\frac{1}{2} \mathrm{ln}(2 \pi {\sigma_{68}}^2) \\
    S_{d}[\mathcal{G}_{2d}] &= 1+\mathrm{ln}(2 \pi {\sigma_{39}}^2)
\end{align}
 for a 1-d Gaussian and symmetric 2-d Gaussian, respectively. The latter is the differential entropy for the approximation of the Fisher-von-Mises distribution for large $\kappa$ 
\begin{align}
\mathrm{FvM}(x;\mu,\kappa) &\stackrel{\phantom{\mathrm{\approx,(3.14)}}}{=} \frac{\kappa \cdot \mathrm{exp}\left(\kappa \cdot \bm{\vec{\mu}}^T  \bm{\vec{x}} \right) }{2 \pi \cdot (e^\kappa - e^{-\kappa})} \ \ \stackrel{\kappa \to \infty}{\rightarrow} \frac{\kappa \cdot \mathrm{exp}\left( \kappa \cdot \mathrm{cos}(\alpha)\right)}{2 \pi \cdot e^\kappa} \\
 &\stackrel{\alpha \to 0}{\mathmakebox[\widthof{$\stackrel{\mathrm{\approx,(5.14)}}{\approx}$}]{\rightarrow}} \frac{\kappa}{2\pi } \cdot\mathrm{exp}\left(- \frac{1}{2} \kappa \alpha^2 \right) \   \ = \frac{1}{2\pi {\sigma_{39} }^2} \cdot\mathrm{exp}\left(- \frac{1}{2} \frac{\alpha^2}{ {\sigma_{39} }^2} \right),
\end{align}
where $\alpha$ is the angle between $\bm{\vec{\mu}}$ and $\bm{\vec{x}}$ and $\sigma_{39}$ is the standard deviation of the 2-d Gaussian approximation for small angles. In the end one obtains a simple relation between the $\kappa$ parameter of the Fisher-von-Mises distribution and $\sigma_{39}$. The Gaussian is a reasonable approximation for $\kappa \gtrsim 30$ which corresponds to $\sigma_{39} \lesssim 10^\circ$. We make use of these relations later to plot the equivalent $\sigma_{68}$ (1-d) and $\sigma_{39}$ (2-d) values for a given entropy, for both energy and direction, respectively.
In order to gauge how non-Gaussian (or for the sphere how non-symmetrical) a particular distribution is, we calculate the KL-divergence between the distribution and its respective maximum entropy approximation with similar first and second order moments. The forward KL-divergence is defined as
\begin{equation}
\mathcal{D}_{\mathrm{KL}}(p(x)|p_{\mathrm{approx.}}(x)) = \int p(x)  \mathrm{log}\left(\frac{p(x)}{p_{\mathrm{approx.}}(x)}\right)dx,
\end{equation}
which is a distance measure between distributions. For the energy distribution, the maximum entropy approximation is a Gaussian distribution. For the directional distribution, we use a FvM distribution, which is a standard assumption in Icecube analyses. Those maximum-entropy approximations are fitted to 10000 samples drawn from the respective normalizing flow. If the normalizing flow is very asymmetric, it will have a large "distance" to its corresponding maximum entropy approximation.
Since normalizing flows allow to produce samples, we can calculate both the entropy and the KL divergence as sampled-based expectation values for each event.

\begin{figure*}[hbt]
\centering
\includegraphics[width=0.46\textwidth]{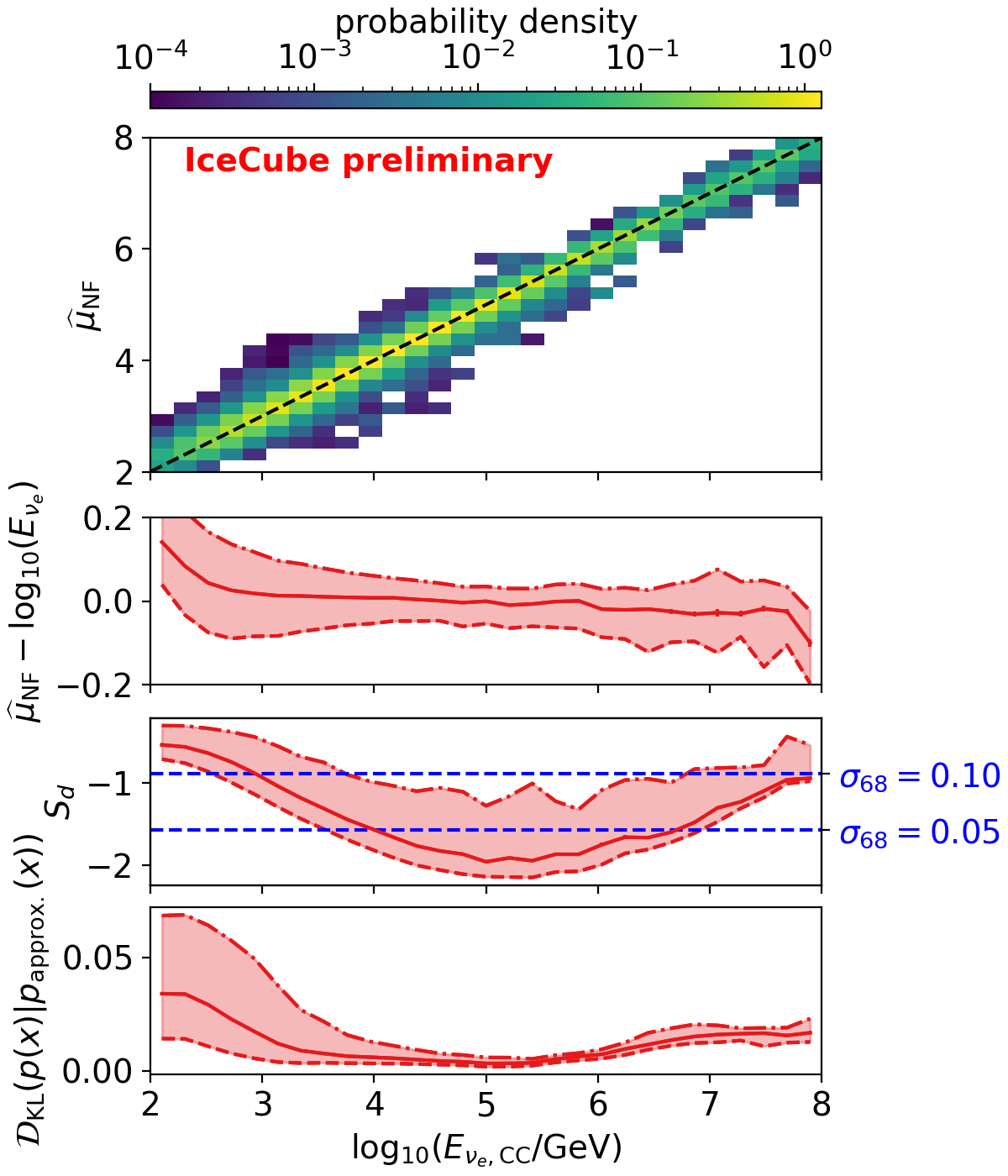}
\includegraphics[width=0.46\textwidth]{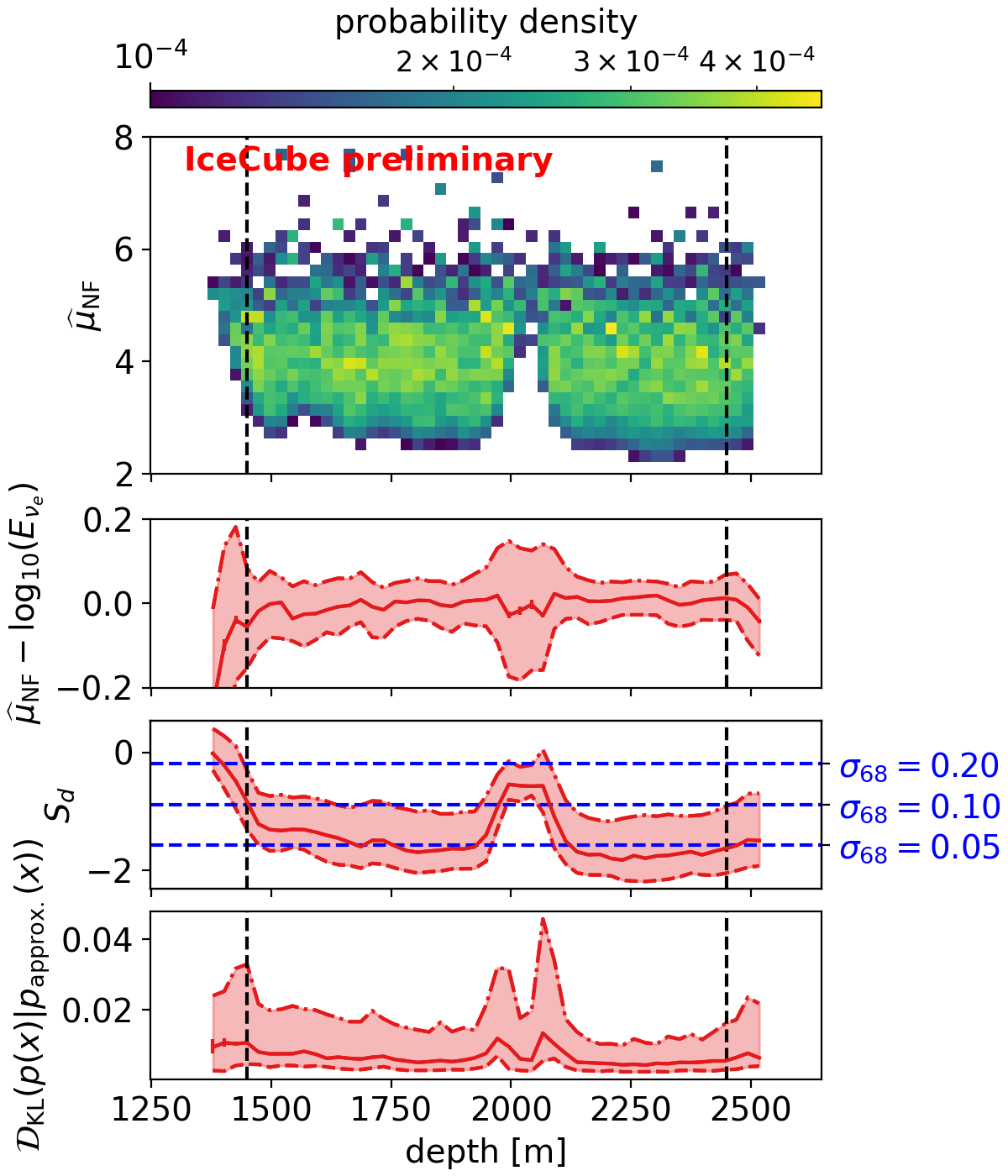}

\caption{Sample mean $\widehat{\mu}_{\mathrm{NF}}$ of the normalizing-flow distribution over $\mathrm{log}_{10}(E_{\nu, \mathrm{CC}})$ versus $\mathrm{log}_{10}(E_{\nu, \mathrm{CC}})$ (left) and depth below the surface (right). The second, third, and fourth row show 16 (dashed), 50 (solid) and 84 (dash-dotted) percentiles of the difference of the true value from the mean, the differential entropy, and the KL-divergence, respectively. Indicated in blue are $\sigma_{68}$ values of the corresponding Gaussian with equivalent entropy to gauge the entropy values. All events are weighted with an $E^{-1.8}$ energy spectrum. The detector boundary is indicated by the dashed black vertical lines.}
\label{fig:energy_resolution}
\end{figure*}

\textbf{\textit{Energy reconstruction of $\nu_{e,CC}$ events:}}
Results for the energy reconstruction are shown in fig.~\ref{fig:energy_resolution}, which depicts the mean of the distribution versus the true energy (left), and versus the depth of interaction (right). The difference between the mean of the normalizing-flow distribution and the true value is centered around zero, with some exception towards the borders of the training dataset range. The spread is smallest in the region around one hundred TeV, and large inside the dust layer, and both of them are reflected in the differential entropy being small and large, respectively. This self-consistency has to happen, in order for coverage (fig \ref{fig:coverage}) to hold. The KL divergence shows that towards smaller energies, in particular, the distributions become less Gaussian. Towards larger energies above $\approx 1 \mathrm{PeV}$ the decrease in resolution is likely due to low training statistics, although it could also in principle be due to PMT saturation effects becoming stronger. This will be looked at in some future study.

\textbf{\textit{Directional reconstruction of $\nu_{e,CC}$ events:}}
\label{sec:electron_dir}
\begin{figure*}
\centering
\includegraphics[width=0.48\textwidth]{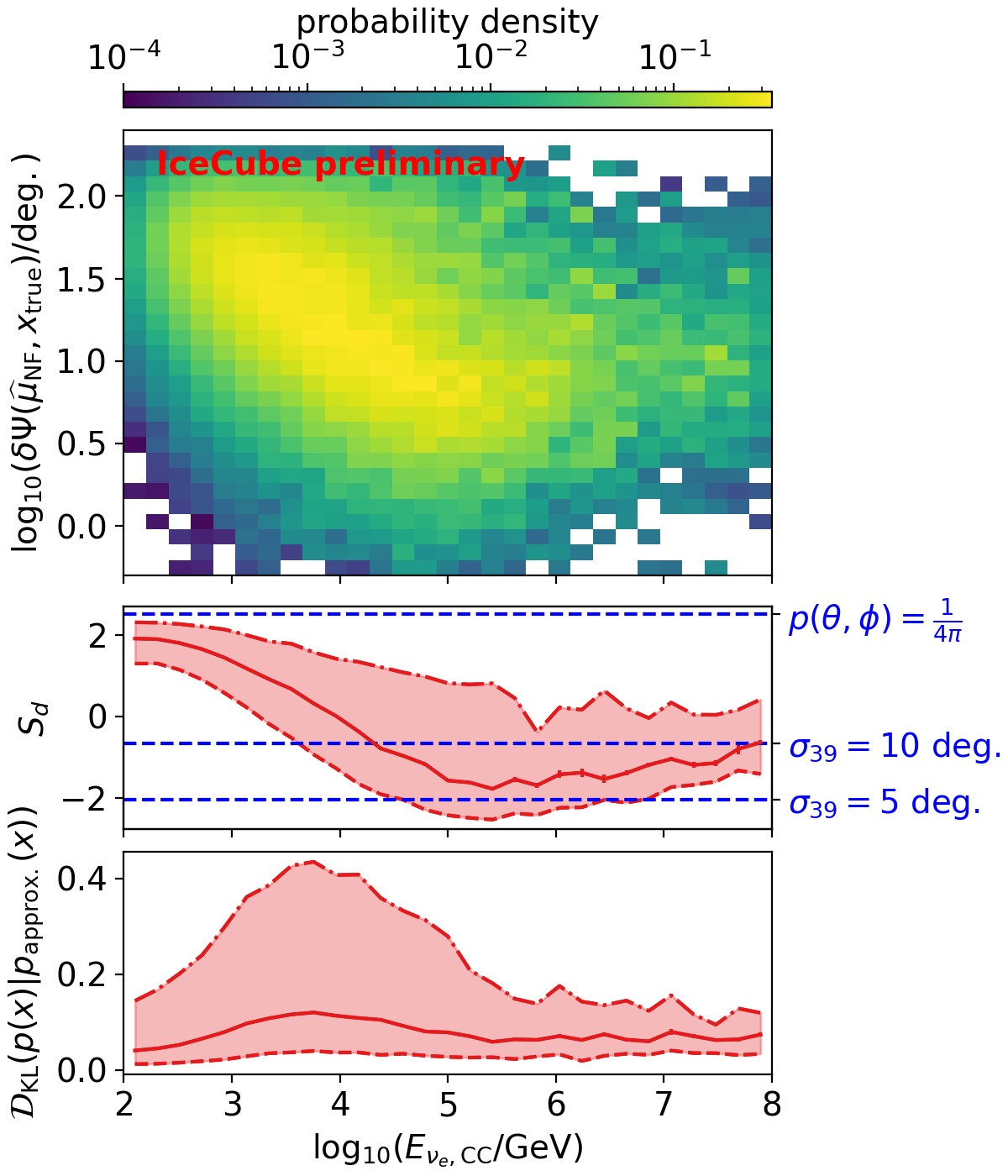}
\includegraphics[width=0.48\textwidth]{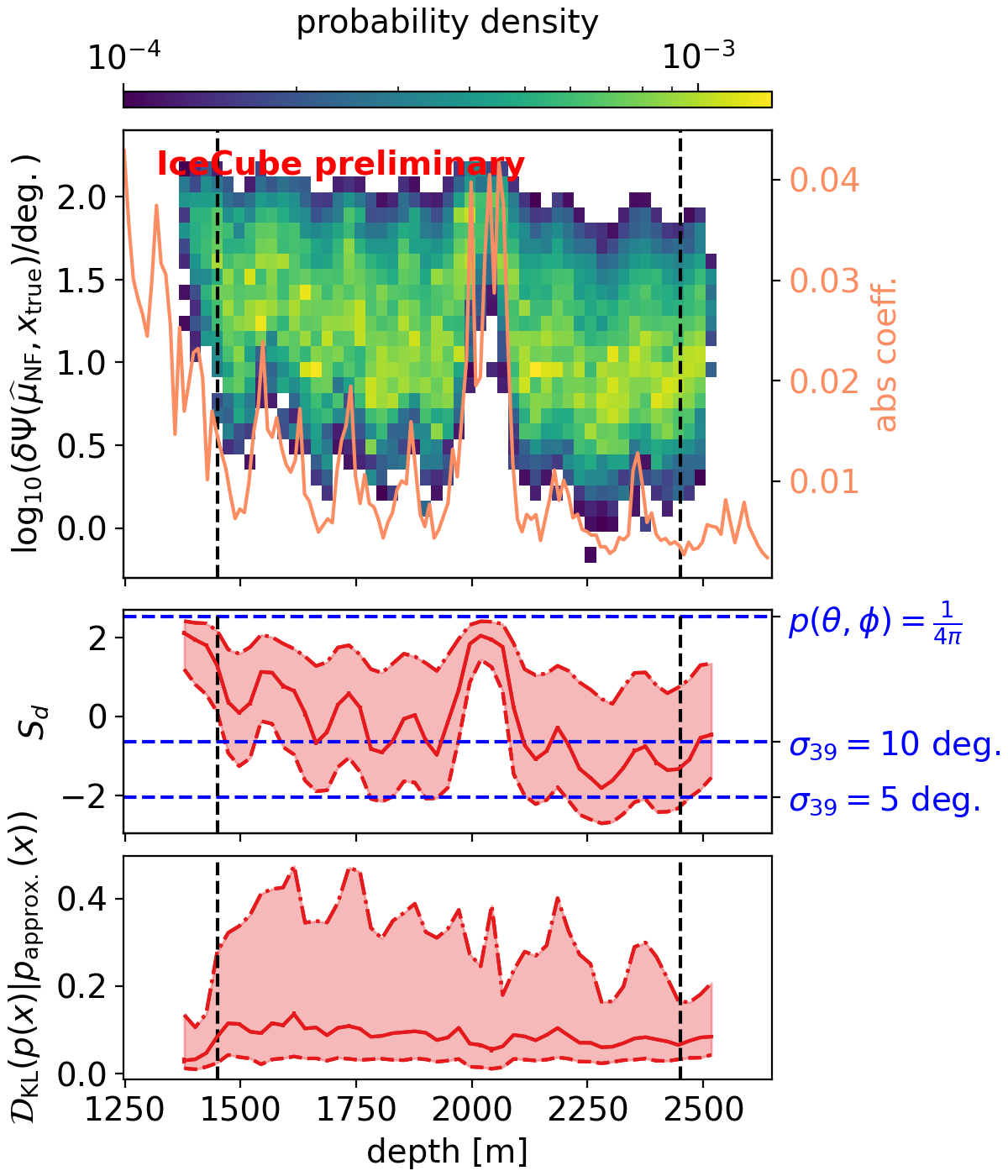}
\caption{Logarithm of the angular distance between sample mean $\widehat{\mu}_{\mathrm{NF}}$ of the normalizing-flow distribution over the direction versus energy (left) and depth below the surface (right). The second and third row show 16 (dashed), 50 (solid) and 84 (dash-dotted) percentiles of the differential entropy, and the KL-divergence, respectively. Indicated in blue are values of $\sigma_{39}$ of a 2-d Gaussian with equivalent entropy and the maximum entropy which corresponds to a flat distribution. All events are weighted with an $E^{-1.8}$ energy spectrum. The detector boundary is indicated by the dashed black vertical lines.}
\label{fig:cascade_directional_reco}
\end{figure*}
Results for the directional reconstruction are shown in fig.~\ref{fig:cascade_directional_reco} which depicts the logarithm of the angular resolution calculated with the mean $\mu$ of the distributions. The left figure shows the angular resolution as a function of energy. It shows that the posterior size reaches the equivalent of a $5-10$ degree 2-d Gaussian around $100 \ \mathrm{TeV}$, which is close to existing shower reconstructions \cite{event_generator_paper}. The KL-divergence indicates a class of events between 1 to 100 TeV that are more asymmetrical. Visual inspection (see fig. \ref{fig:high_kl_example}) shows that these are events with large azimuthal uncertainty. Symmetric reconstructions based on a FvM distribution would fail to describe such events and overpredict the uncertainty. The recent analysis of the galactic plane \cite{galactic_plane_results} used the FvM assumption, and therefore could potentially be improved with normalizing flows. The right figure depicts the resolution as a function depth below the surface. The correlation to optical ice properties is clearly visible, and demonstrates that complex behavior is automatically learned by the normalizing flow. 

\begin{figure*}
\centering
\includegraphics[width=0.49\textwidth]{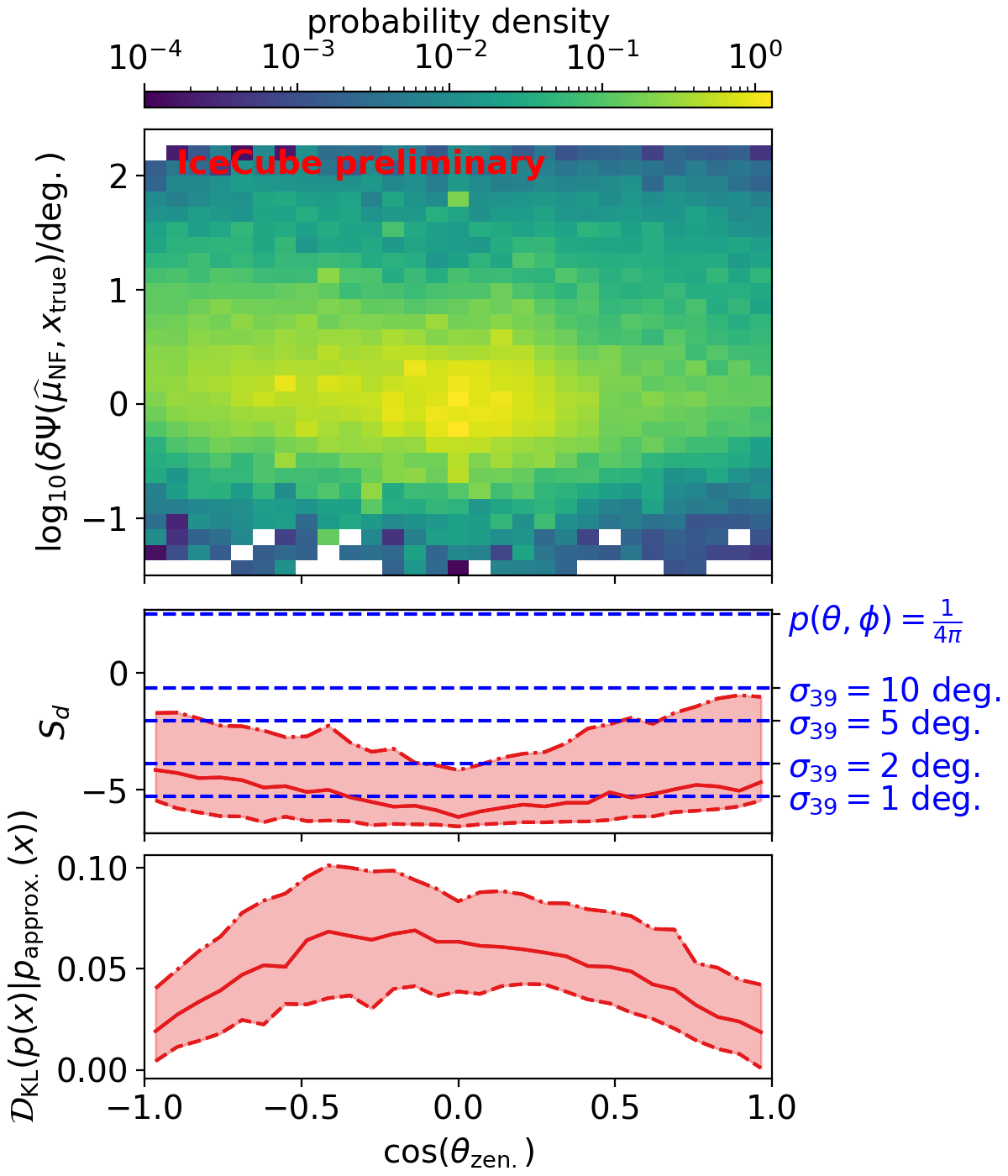}
\includegraphics[width=0.49\textwidth]{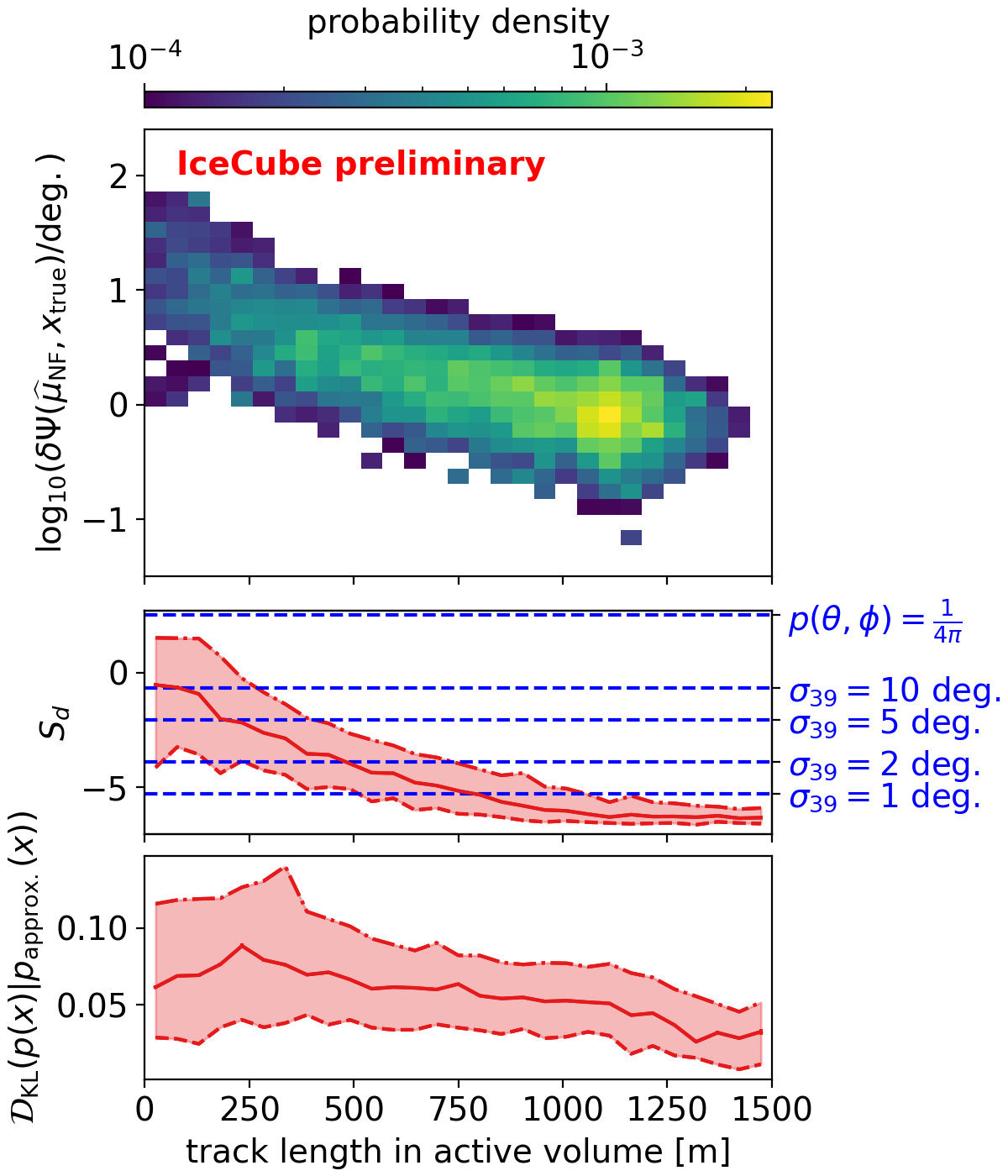}
\caption{Logarithm of the angular distance between sample mean $\widehat{\mu}_{\mathrm{NF}}$ of the normalizing-flow distribution over the muon direction versus the cosine of the zenith direction (left) and muon length within the active detector volume (right). The second and third row show 16 (dashed), 50 (solid) and 84 (dash-dotted) percentiles of the differential entropy, and the KL-divergence, respectively. Indicated in blue are values of $\sigma_{39}$ of a 2-d Gaussian with equivalent entropy and the maximum entropy which corresponds to a flat distribution. All events are weighted with an $E^{-1.8}$ energy spectrum.}
\label{fig:muon_reconstruction}
\end{figure*}

\textbf{\textit{Directional reconstruction of $\nu_{\mu,CC}$ events:}}
Figure \ref{fig:muon_reconstruction} depicts results for the spherical normalizing flow for muons. The logarithm of the angular resolution calculated with the mean $\mu$ of the distributions versus the cosine of the incident neutrino zenith is shown on the left. The performance is highest at the equator, with resolutions below one degree, and worse for up and downgoing directions. On the right of the figure, which shows the resolution as a function of contained track length, one can see a correlation of the performance with larger track lengths, as expected. While this performance likely does not match existing muon reconstructions yet, other encoding strategies that incorporate more label information for muons can achieve the performance of leading likelihood approaches, as was shown in \cite{rnn_moon}. The KL-divergence, which measures the asymmetry of contours, tends to be larger for more horizontal muon tracks and for tracks that are only partially contained. The asymmetry, however, is overall smaller than for electron neutrino showers.

\section{Conclusion}
In this contribution we trained conditional normalizing flows to learn the per-event posterior distributions over (log-)energy and direction of charged-current electron- and muon neutrinos in IceCube. The learned posterior distributions $p(\theta;x)$ directly use a summary statistic of the high-dimensional noise-cleaned photon data $x$ as input. Coverage of these posteriors can be verified without numerical scans and is within a few percent of the expected values. For horizontal muon tracks, the median angular resolution of the normalizing-flow mean reaches below 1 degree. For electron neutrino showers, the angular normalizing flow approaches a performance of 5-10 degrees median angular resolution of the normalizing-flow mean around 100 TeV, which is comparable to state of the art likelihood approaches. We additionally find that between 1 and 100 TeV many electron neutrino events have large asymmetric uncertainties where the azimuth is much less constrained. For track reconstructions, the asymmetries tend to be smaller, but still detectable. In both cases, these asymmetric contours show the potential to improve on existing analyses, which currently utilize symmetric contour assumptions.

Further interesting applications that are encouraged by these results are to use them for very low-energy events in oscillation analyses whose contours are expected to be non-Gaussian or real-time alert PDFs which require coverage guarantees.

\bibliographystyle{ICRC}
\setlength{\bibsep}{1.pt}
\bibliography{main}

\providecommand{\href}[2]{#2}\begingroup\raggedright\begin{thebibliography}{10}

\bibitem{hese_refined}
{The IceCube Collaboration}
  \href{http://dx.doi.org/10.1103/PhysRevD.104.022002}{{\em Physical Review
  Letters} {\bfseries 104} no.~2, (July, 2021) 022002}.

\bibitem{txs}
{The IceCube Collaboration}
  \href{http://dx.doi.org/10.1126/science.aat2890}{{\em Science} {\bfseries
  361} no.~6398, (2018) 147--151}.

\bibitem{ps_improvements}
{The IceCube Collaboration}
  \href{http://dx.doi.org/10.1126/science.abg3395}{{\em Science} {\bfseries
  378} no.~6619, (Nov., 2022) 538--543}.

\bibitem{cov_paper}
T.~{Gl{\"u}senkamp} \href{http://dx.doi.org/10.48550/arXiv.2008.05825}{{\em
  arXiv e-prints} (Aug., 2020) arXiv:2008.05825}.

\bibitem{nf_spheres}
D.~J. Rezende {\em et~al.}, ``Normalizing flows on tori and spheres,'' in {\em
  ICML}, pp.~8083--8092.
\newblock 2020.

\bibitem{gaussianization_flows}
C.~{Meng}, Y.~{Song}, J.~{Song}, and S.~{Ermon}
  \href{http://dx.doi.org/10.48550/arXiv.2003.01941}{{\em arXiv e-prints}
  (Mar., 2020) arXiv:2003.01941}.

\bibitem{jf}
T.~Gl\"usenkamp  (2023) \texttt{https://github.com/thoglu/jammy\_flows}.

\bibitem{gnn_paper}
{The IceCube Collaboration}
  \href{http://dx.doi.org/10.1088/1748-0221/17/11/P11003}{{\em Journal of
  Instrumentation} {\bfseries 17} no.~11, (Nov, 2022) P11003}.

\bibitem{event_generator_paper}
{The IceCube Collaboration} \href{http://dx.doi.org/10.22323/1.395.1065}{{\em
  PoS} {\bfseries 395} (Mar., 2022) 1065}.

\bibitem{galactic_plane_results}
{The IceCube Collaboration}
  \href{http://dx.doi.org/10.1126/science.adc9818}{{\em Science} {\bfseries
  380} no.~6652, (2023) 1338--1343}.

\bibitem{rnn_moon}
{The IceCube Collaboration} \href{http://dx.doi.org/10.22323/1.395.1087}{{\em
  PoS} {\bfseries 395} (Mar., 2022) 1087}.

\end{thebibliography}\endgroup

\clearpage

\section*{Full Author List: IceCube Collaboration}

\scriptsize
\noindent
R. Abbasi$^{17}$,
M. Ackermann$^{63}$,
J. Adams$^{18}$,
S. K. Agarwalla$^{40,\: 64}$,
J. A. Aguilar$^{12}$,
M. Ahlers$^{22}$,
J.M. Alameddine$^{23}$,
N. M. Amin$^{44}$,
K. Andeen$^{42}$,
G. Anton$^{26}$
% ...
\\
\\
% ...
$^{12}$ Universit{\'e} Libre de Bruxelles, Science Faculty CP230, B-1050 Brussels, Belgium \\
% ...
$^{17}$ Department of Physics, Loyola University Chicago, Chicago, IL 60660, USA \\
$^{18}$ Dept. of Physics and Astronomy, University of Canterbury, Private Bag 4800, Christchurch, New Zealand \\
% ...
$^{22}$ Niels Bohr Institute, University of Copenhagen, DK-2100 Copenhagen, Denmark \\
$^{23}$ Dept. of Physics, TU Dortmund University, D-44221 Dortmund, Germany \\
% ...
$^{26}$ Erlangen Centre for Astroparticle Physics, Friedrich-Alexander-Universit{\"a}t Erlangen-N{\"u}rnberg, D-91058 Erlangen, Germany \\
% ...
$^{40}$ Dept. of Physics and Wisconsin IceCube Particle Astrophysics Center, University of Wisconsin{\textendash}Madison, Madison, WI 53706, USA \\
% ...
$^{42}$ Department of Physics, Marquette University, Milwaukee, WI, 53201, USA \\
% ...
$^{44}$ Bartol Research Institute and Dept. of Physics and Astronomy, University of Delaware, Newark, DE 19716, USA \\
% ...
$^{63}$ Deutsches Elektronen-Synchrotron DESY, Platanenallee 6, 15738 Zeuthen, Germany  \\
$^{64}$ Institute of Physics, Sachivalaya Marg, Sainik School Post, Bhubaneswar 751005, India \\
% ...

\subsection*{Acknowledgements}

\noindent
The authors gratefully acknowledge the support from the following agencies and institutions:
USA {\textendash} U.S. National Science Foundation-Office of Polar Programs,
U.S. National Science Foundation-Physics Division,
U.S. National Science Foundation-EPSCoR,
Wisconsin Alumni Research Foundation,
Center for High Throughput Computing (CHTC) at the University of Wisconsin{\textendash}Madison,
Open Science Grid (OSG),
Advanced Cyberinfrastructure Coordination Ecosystem: Services {\&} Support (ACCESS),
Frontera computing project at the Texas Advanced Computing Center,
U.S. Department of Energy-National Energy Research Scientific Computing Center,
Particle astrophysics research computing center at the University of Maryland,
Institute for Cyber-Enabled Research at Michigan State University,
and Astroparticle physics computational facility at Marquette University;
Belgium {\textendash} Funds for Scientific Research (FRS-FNRS and FWO),
FWO Odysseus and Big Science programmes,
and Belgian Federal Science Policy Office (Belspo);
Germany {\textendash} Bundesministerium f{\"u}r Bildung und Forschung (BMBF),
Deutsche Forschungsgemeinschaft (DFG),
Helmholtz Alliance for Astroparticle Physics (HAP),
Initiative and Networking Fund of the Helmholtz Association,
Deutsches Elektronen Synchrotron (DESY),
and High Performance Computing cluster of the RWTH Aachen;
Sweden {\textendash} Swedish Research Council,
Swedish Polar Research Secretariat,
Swedish National Infrastructure for Computing (SNIC),
and Knut and Alice Wallenberg Foundation;
European Union {\textendash} EGI Advanced Computing for research;
Australia {\textendash} Australian Research Council;
Canada {\textendash} Natural Sciences and Engineering Research Council of Canada,
Calcul Qu{\'e}bec, Compute Ontario, Canada Foundation for Innovation, WestGrid, and Compute Canada;
Denmark {\textendash} Villum Fonden, Carlsberg Foundation, and European Commission;
New Zealand {\textendash} Marsden Fund;
Japan {\textendash} Japan Society for Promotion of Science (JSPS)
and Institute for Global Prominent Research (IGPR) of Chiba University;
Korea {\textendash} National Research Foundation of Korea (NRF);
Switzerland {\textendash} Swiss National Science Foundation (SNSF);
United Kingdom {\textendash} Department of Physics, University of Oxford.

\end{document}